\documentclass[11pt]{article}
\usepackage{amsmath,amsfonts}
\usepackage{amssymb}
\usepackage{fullpage}
\usepackage{lscape}
\usepackage[dvips]{epsfig}

\def\##1{\underline #1}
\def\=#1{\underline{\underline #1}}
\def\.{\mbox{ \tiny{$^\bullet$} }}
\def\diffz{\frac{\rm d}{{\rm d} z}}
\def\curl{\nabla\times}

\def\muo{\mu_0}
\def\eps{\epsilon}
\def\epso{\eps_0}

\def\ux{{\bf u}_x}
\def\uy{{\bf u}_y}
\def\uz{{\bf u}_z}

\def\Einc{{\bf E}_{\rm i}}
\def\Hinc{{\bf H}_{\rm i}}

\def\FDE{\tilde{\bf E} \le {\bf r}, \omega  \ri}
\def\FDH{\tilde{\bf H} \le {\bf r}, \omega \ri}
\def\TDE{{\bf E} \le {\bf r}, t  \ri}
\def\TDH{{\bf H} \le {\bf r}, t \ri}
\def\FDEc{\tilde{\bf E}^\ast \le {\bf r}, \omega  \ri}

\def\mf{\les \tilde{\bf \#f} \le z, \omega \ri \ris}

\def\mP{\les \tilde{\bf \=P} \le z, \omega \ri \ris}

\def\SAR{{\rm SAR} ({\bf r})}

\def\le{\left(}
\def\ri{\right)}
\def\les{\left[}
\def\ris{\right]}

\def\Nt{N_{\rm t}}

\def\c#1{\cite{#1}}

\def\r#1{(\ref{#1})}
\bigskip
\begin{document}

%\begin{center}

\bigskip
\noindent{\Large {\bf Electromagnetic modeling of
 near--field phase--shifting contact lithography with broadband ultraviolet illumination  } }
\vskip 0.4cm

\noindent {\bf Fei Wang$^a$},
 {\bf Katherine E. Weaver$^a$},
 {\bf Akhlesh Lakhtakia$^{a,b}$}\footnote{Corresponding
Author. Tel: +1-814-863-4319, Fax: +1-814-865-9974, e-mail:
akhlesh@psu.edu} and {\bf Mark W. Horn$^a$}

\vskip 0.2cm
\noindent $^a${\em CATMAS~---~Computational \& Theoretical
Materials Sciences
Group,
Department of Engineering Science \& Mechanics,
Pennsylvania State University, University Park, PA 16802--6812,
USA}

\vskip 0.2cm
\noindent $^b${\em Photonics Section, Department of Physics,
Imperial College London,
London SW7 2BZ, UK}

%\end{center}
\bigskip

 \noindent {\bf Abstract.}  Near--field phase--shifting contact lithography
 is modeled to characterize electromagnetic absorption
in a photoresist layer with one face in contact
with a quartz binary phase--shift mask. The broadband ultraviolet illumination
 is represented as a
frequency--spectrum of normally incident plane waves. A
rigorous coupled--wave analysis is carried out to determine the
absorption  spectrum of the photoresist layer. The specific absorption rate
in the photoresist layer is calculated and examined in relation to
the geometric parameters. Columnar features  in the photoresist
layer are of higher quality on broadband illumination in contrast to
monochromatic illumination, in conformity with some recent experimental results.  Feature resolution
and profile are noticeably affected by the depth of the grooves in the
phase--shift mask. Ideally, the feature linewidth can be less than about 100 nm
for broadband illumination in the transverse--magnetic mode. These conclusions
are subject to modification by the  photochemistry--wavelength characteristics of the photoresist.


\section{Introduction}
Several near--field imaging techniques yielding sub--wavelength
resolution have been recently reported \c{1}--\c{7}. In
particular,  {\em near-field phase--shifting contact lithography}
has been demonstrated to extend the resolution of contact aligners
to less than $200$~nm by using single--layer photoresists with
high aspect ratios \c{7, 8}. The aerial image generated by NFPSCL
was suggested to be due to a combination of near--field and
phase--shifting effects  \c{4}. But features are resolved
uniformly throughout photoresist layers of thicknesses far in
excess of the exposure wavelength, which means that the dominance
of near--field effects is doubtful and may even be absent
 \c{JP1}--\c{JP3}. In contrast, a recent
experimental investigation on broadband ultraviolet (UV) lithography
 disclosed the high sensitivity of the feature
resolution  to the phase shifts built into the phase mask
\c{8}. That phase--shifting effects are crucial to the success of NFPSCL
while near--field effects are not always significant, was
also confirmed by numerical simulation \c{MEE}.
But one question still remains: why are the smallest linewidths
achieved through NFPSCL with the use of broadband UV
illumination, rather than with monochromatic illumination, particularly because of degradation of
phase--shifting effects due to the presence of a wide spectrum in broadband
illumination?

Electromagnetic modeling of NFPSCL has been carried out
with different numerical methods by several researchers with
different objectives. Aizenberg {\em et al.\/}
\c{5} presented a simple model for the near--field effect but
did not account for the phase--shifting effect.  Kunz {\em et
al.\/} \c{9} presented the finite--difference--time--domain (FDTD)
modeling of NFPSCL on flexible substrates, focusing on
top--surface imaging photoresists. More recently, we used the rigorous coupled--wave analysis (RCWA) to model
electromagnetic absorption in the photoresist layer for
monochromatic UV illumination \c{MEE}. We found that
columnar features are transversely localized in the photoresist layer close to the
edges of the periodically corrugated mask, as a result of the superposition of
propagating Floquet harmonics. The evanescent Floquet harmonics
play no role in this spatial localization.

The localization of
absorption can be
enhanced further by means of the  superposition of
many sets of propagating Floquet harmonics vibrating at different frequencies. In fact,
preliminary modeling indicated  noticeable
improvement in feature resolution and profile on replacing
monochromatic illumination by incoherent trichromatic illumination
\c{MEE}.

Motivated by those theoretical findings as well as by experimental
data \c{8}, we undertook the broadband--illumination modeling of
NFPSCL, even though the commonplace industrial practice is to use
quasi\-mono\-chromatic (i.e., narrowband) illumination. Our results are reported here. As part of our technique,
the electromagnetic field of the broadband source of illumination
is represented by a superposition of plane waves of different
frequencies and different wavevectors. The RCWA is performed  to
the variation of absorption with frequency at any location in the
photoresist layer. The overall specific absorption rate (SAR) is
then calculated to characterize the columnar features in the
photoresist layer, which are certainly the precursors of aerial
images obtained after development.

A significant conclusion is that broadband UV illumination can produce acceptable
results, in contrast to the undesirable standing--wave patterns in the photoresist layer produced by monochromatic UV
illumination. Even though the use of antireflection coatings (ARCs) can drastically
reduce the standing--wave patterns, broadband illumination does not require the
additional step of putting on an ARC; furthermore, filtering optics is also not needed with
broadband illumination. We caution, however, that our electromagnetic modeling requires coupling
with the spectral characteristics of the photochemistry of the photoresist, which coupling  lies outside the
scope of this paper; nevertheless, experimental results \c{8} are consistent with our conclusion.

\section{Theoretical Analysis}

The electromagnetic boundary value problem is schematically shown
in Fig. 1. The three regions $0 < z < h_1$, $h_2<z<h_3$, and
$h_3<z<h_4$ are occupied, respectively, by homogeneous materials
labeled $a$, $c$, and $d$; and the corresponding relative
permittivity scalars are denoted by $\eps_a$, $\eps_c$, and
$\eps_d$. The region $h_1<z<h_2$ acts as a binary phase--shift
mask with alternate strips of widths $qL$ and $(1-q)L$, $0\leq q
\leq 1$, made of materials labeled $a$ and $b$. The half--spaces
$z\leq 0$ and $z\geq h_{4}$ are vacuous.
 Material $a$ is quartz, material $b$ is air (equivalently, vacuum),
 material $c$ is the chosen photoresist, while material $d$ is
 silicon.
 For convenience, we define the  thicknesses
 $\Delta h_j=h_j-h_{j-1}$, $j\in [1, \,4]$, where
 $h_0=0$.

Broadband  light is incident from the half--space $z\leq 0$ on
to the plane $z=0$. As a result, reflection and transmission into
the two half--spaces, $z\leq 0$ and $z\geq h_4$, respectively, occur. The incident electromagnetic field is
represented in the time--domain through the temporal Fourier transform as
\begin{equation}
\label{E1} \Einc ({ \bf r}, t)=\int\limits_{-\infty}^\infty
\tilde{\bf E}_{\rm i} ({\bf r}, \omega) {\rm e}^{-i \omega t} {\rm
d} \omega \,,  \quad  \Hinc ({ \bf r},
t)=\int\limits_{-\infty}^\infty \tilde{\bf H}_{\rm i} ({\bf r},
\omega) {\rm e}^{-i \omega t} {\rm d} \omega \,,
\end{equation}
where ${\bf r}=x\ux+y\uy+z\uz$ and $t$ represent the position
vector and time, respectively, $\omega$ is the angular frequency,
and  $i=\sqrt{-1}$.

For compatibility with commonplace industrial
usage, the field phasors $\tilde{\bf E}_{\rm i} ({\bf r}, \omega)$
and $\tilde{\bf H}_{\rm i} ({\bf r}, \omega)$ at any $\omega$ are
taken to be associated with a plane wave propagating in the $+z$
direction; therefore,
\begin{eqnarray}
\label{E2}
 \tilde{\bf E}_{\rm i} ({\bf r}, \omega) &=& A(\omega) \le   a_s\,\uy -
 a_p\,\ux \ri {\rm exp} ( ik_0z) \,, \\
 \tilde{\bf H}_{\rm i} ({\bf r}, \omega) &=& -\,\frac{1}{\eta_0} A(\omega) \le
a_s\,\ux +   a_p\,\uy \ri {\rm exp}(ik_0z) \,, \label{E3}
\end{eqnarray}
where  $I(\omega)=A^2(\omega)$ is the incident light's spectral
intensity function, $\eta_0=\sqrt{\muo/\epso}$ is the intrinsic
impedance of vacuum,  $k_0 =\omega \sqrt{\muo
\epso}=2\pi/\lambda_0$ is the vacuum wavenumber, and  $\lambda_0$
is the wavelength in vacuum, $\muo$ is the permeability of vacuum, and
$\epso$ is the permittivity of vacuum. The amplitudes $a_s$ and $a_p$
determine the linear polarization state of the incident
electromagnetic field, and are subject to the condition
\begin{equation}
a_s^2+  a_p^2=1 \,.
\end{equation}

As the total
electromagnetic field everywhere is represented as
\begin{equation}
\label{E4} {\bf E} ({ \bf r}, t)=\int\limits_{-\infty}^\infty
\tilde{\bf E} ({\bf r}, \omega) {\rm e}^{-i \omega t} {\rm d}
\omega \,,  \quad  {\bf H} ({ \bf r},
t)=\int\limits_{-\infty}^\infty \tilde{\bf H} ({\bf r}, \omega)
{\rm e}^{-i \omega t} {\rm d} \omega \,,
\end{equation}
our next task is to calculate the field phasors $\tilde{\bf E}
({\bf r}, \omega)$ and $\tilde{\bf H} ({\bf r}, \omega)$ in terms
of $\tilde{\bf E}_{\rm i} ({\bf r}, \omega)$ and $\tilde{\bf
H}_{\rm i} ({\bf r}, \omega)$ for arbitrary $\omega$ and ${\bf
r}$. Because of the planewave format of $\tilde{\bf E}_{\rm i}
({\bf r}, \omega)$ and $\tilde{\bf H}_{\rm i} ({\bf r}, \omega)$,
the method of choice is RCWA.

Detailed accounts of RCWA are commonplace in the optics literature \c{GM,BJbook}. As a complete
account for the problem described via Figure 1 has been presented by us
elsewhere \c{MEE}, we just reproduce here the essence of the technique. Let the relative permittivity scalar
be denoted by $\epsilon({\bf r},\omega)$. Because of
\begin{itemize}
\item[(i)] the
$x$--periodicity of $\eps ({\bf r},\omega)$ for $z \in (h_1, h_2)$, and
\item[(ii)] the uniformity of $\eps ({\bf r},\omega)$  $\forall z\in(-\infty,\infty)$ along the
$y$ axis,
\end{itemize}
the total
field phasors $\tilde{\bf E} ({\bf r}, \omega)$ and $\tilde{\bf H}
({\bf r}, \omega)$ can be decomposed everywhere in terms of
Floquet harmonics as follows:
\begin{eqnarray}
\label{2.17} \FDE&=&\sum_{n\in \mathbb{Z}} \,\tilde{\bf E}^{(n)}
(z, \omega)  \exp\le in\frac{2\pi x}{L}\ri\,, \\[5pt]
\FDH&=&\sum_{n\in \mathbb{Z}} \,\tilde{\bf H}^{(n)} (z, \omega)
\exp \le in\frac{2\pi x}{L}\ri \,. \label{2.18}
\end{eqnarray}
These field phasors  must satisfy the
frequency--domain Maxwell curl postulates everywhere.

Specifically, the equations
\begin{equation}
\label{2.15} \left.
\begin{array}{l}
\curl \FDE=i  \omega\muo\, \FDH \\
\curl \FDH=-i \omega\epso \, \eps ({\bf r},\omega)\, \FDE
\end{array}
\right\}
\end{equation}
hold for $z\in (0, \, h_4)$. After expanding $\eps ({\bf r},\omega)$ into a
Fourier series with respect to $x$   and substituting \r{2.17} and
\r{2.18} into \r{2.15}, the matrix ordinary differential equation
\begin{equation}
\label{2.39} \diffz \mf=i\,\mP  \mf \,
\end{equation}
is derived for $z\in (0, \,h_4)$, where the column vector $\mf$
contains the $x$-- and the $y$--directed components of both $\tilde{\bf E}^{(n)} (z,
\omega)$ and $\tilde{\bf H}^{(n)} (z, \omega)$.

For  digital computation, the  restriction $|n| \leq \Nt$ is necessary.
 Floquet expansions of the reflected field phasors in the half--space $z\leq 0$ and of
 the transmitted field phasors in the half--space $z\geq h_4$ are set up. Equation
 \r{2.39} is then solved, after enforcing the
continuity of  the $x$-- and the $y$--directed
components of the electromagnetic field phasors across the planes
$z=0$ and $z=h_4$. The parameter $\Nt$ has to be increased until a convergent solution of
 \r{2.39} is found \c{MEE}.

Once the total field phasors $\FDE$ and $\FDH$ have been obtained everywhere
by using RCWA, $\TDE$ and $\TDH$
can be determined from \r{E4} for any ${\bf r} $ and
$t$ by using the inverse Fourier transform.
Our interest, however, lies only in the electromagnetic energy absorbed in the
photoresist layer and converted into both thermal
and chemical forms therein. Let the illumination be carried
out only for $t\in[0,\,T]$. The (time--averaged)
specific absorption rate ${\rm SAR} ({\bf r})$ at a point ${\bf
r}$ is quantitated by
\begin{equation}
\label{SAR1} {\rm SAR}({\bf r})= \omega \, \epso\, \mbox{Im}
\les\eps({\bf r},\omega)\ris \frac{1}{T} \int\limits_0^T \TDE \cdot \TDE
{\rm d} t \,,
\end{equation}
for a quasimonochromatic field, with the assumption
of no dispersion. In fact, when the
field is monochromatic, i.e., $\TDE={\rm Re} \les\tilde{\bf E} ({\bf r})
{\rm exp} ({-i \omega t}) \ris$, \r{SAR1} yields the identity
\begin{equation}
{\rm SAR}({\bf r})=\frac{1}{2}\, \omega \, \epso\, \mbox{Im}
\les\eps({\bf r},\omega)\ris |\tilde{\bf E} ({\bf r})|^2\,,
\end{equation}
which is well--known in the electromagnetics literature  \cite[Eq. 7-60]{Johnk}. If the temporal variation
of $\TDE$ is known, for our purposes
\r{SAR1} may provide a proper, though not exact, estimation of
${\rm SAR}({\bf r})$. However, it is advantageous to represent
${\rm SAR}({\bf r})$ as a superposition of spectral dissipative
contributions. In fact, according to Plancherel's theorem \cite[p. 183]{Evans}
\begin{equation}
\int\limits_{-\infty}^\infty \TDE \cdot \TDE {\rm d} t =
\int\limits_{-\infty}^\infty \FDE \cdot \FDEc {\rm d} \omega \,;
\end{equation}
hence, we use the estimate
\begin{equation}
\label{SAR2} {\rm SAR}({\bf r}) \simeq \frac{1}{4\pi}
(\omega_l+\omega_u) \, \epso\, 
\int\limits_{\omega_l}^{\omega_u} \omega\,\mbox{Im} \les\eps({\bf r},\omega)\ris\, \FDE \cdot \FDEc {\rm d}
\omega \,,
\end{equation}
where $\omega_l$ and $\omega_u$ are two extremities of the
frequency--band of $\FDE$.  We implemented the right side of \r{SAR2} in a representative
element (RE) of material $c$, the RE being the $\Delta h_3\times L$ rectangle in Fig. 1.

Absorption of  photonic energy
 and subsequent curing together play a
significant role in the formation of photoresist features after
development. Therefore, the spatial characteristics of ${\rm
SAR}({\bf r})$  provide direct information on
the photoresist features developed~---~which, in other words, indicate pattern transfer from the mask (material $a$)
 to the photoresist layer (material $c$).

\section{\bf Results and Discussion}
For illustrative results, we chose the following representative
materials: material $a$ is quartz with $\eps_a=1.48^2$; material
$b$ is air so that $\eps_b=1.0$; material $c$ is the photoresist
SPR 505 whose refractive index is plotted in Fig. 2(a) for
$\lambda_0 \in [250, \, 610] $ nm \c{SPR}; material $d$ is
crystalline silicon whose refractive index is presented in Fig.
2(b) for $\lambda_0 \in [250, \, 610] $ nm \c{silicon}. In
accordance with our earlier  paper \c{MEE}, the thicknesses
$\Delta h_1=6$ mm, $\Delta h_3=1$ $\mu$m, and $\Delta h_4=1$ mm
were chosen. While the thickness $\Delta h_2=460$ nm of the binary
phase--shift mask was fixed for most calculations, other values of
$\Delta h_2$ were also adopted for comparative studies. Two values
of $L$ ($=3$ and $4$~$\mu$m) and three values of the ratio $q$
($=0.2$, $0.5$ and $0.8$) were chosen.

The broadband UV source was chosen to be of the UV400 type
employed in commercial S\"uss Microtec mask aligners \c{Suss}. The
spectral intensity function ${ I} (\lambda_0)$ of this source is
plotted in Fig. 3 for $\lambda_0 \in [250,\, 610]$ nm. This
spectral regime was uniformly discretized into 136 subregimes, in
each of which the electric field phasor $\FDE$ was represented
with the value calculated at the central frequency   of the
subregime. Calculations of $\FDE$ were carried out after
ascertaining that $\Nt=12$  sufficed to yield convergent results
for all $\lambda_0 \in [250,\, 610]$ nm. As the incident light can
be linearly linearly polarized, we set $a_s=1$ for transverse
electric (TE) fields  and $a_p=1$ for transverse magnetic (TM)
fields, respectively.\footnote{The electric (resp. magnetic) field
of a TE (resp. TM) field does not have $x$-- and $z$--directed
components, and is thus directed parallel to the grooves of the
phase--shift mask.} We ensured that the principle of energy
conservation was not violated  by any of the results reported here
\c{MEE}.

Figs. 4--6 show gray--level (black implies low magnitudes, white
implies high) plots of $\SAR$ throughout the representative
element of the photoresist layer (see Fig. 1)  for the three
different values of $q$,  when the illumination is broadband.
Figs. 7--9 present the analogous $\SAR$ plots, but for
monochromatic illumination of wavelength
$\lambda_0=2(\sqrt{\eps_a}-1) \Delta h_2= 441 $ nm. In all of
these figures, the letter ``Q'' specifies an $x$--axis range of
${\bf r}$ that is right underneath the strip of material $a$
(quartz), while ``A'' specifies the range of ${\bf r}$ underneath
the strip of material of $b$ (air). Results for both TE and TM illumination
modes are presented in Figs. 4--9.

Clearly, the plots for the TE and TM illumination modes in Figs. 4--9 look quite
different. The most prominent feature of Figs.
4--6, as compared with Figs. 7--9, is the enhanced localization of
power dissipation, and the resulting improvement of black--colored
columnar features. These columnar features denote those portions
of the photoresist layer in which little electromagnetic energy is
dissipated
 and therefore remain on the
substrate after the development process. Our results indicate that, by using broadband UV
light sources, the columnar features are dramatically localized
underneath the vicinity of the phase edges (intersections) of the
Q and A strips, while power dissipation occurs quite homogeneously
outside the columnar features. In contrast, when the incident
light is monochromatic, the columnar features are not
resolved as nicely, which can be deduced
from the presence of black--colored transverse
strips in Figs. 7--9. The appearance of these strips is due to the
standing--wave characteristic of the monochromatic field in the
photoresist layer \c{MEE}.

Undoubtedly, our theoretical results indicate that better feature
resolution and profile would be achieved by using broadband
illumination in place of monochromatic illumination. This
conclusion coincides with the implications of recent experiments
\c{8}. High--aspect--ratio columnar features are predicted by the
model, especially for TM illumination, with linewidths less than
$200$ nm and profiles quite uniform on one side~---~as shown in
Fig. 5(b).

Furthermore, the ratio $q$ affects the position as well
as the profile/resolution of the columnar features, which
conclusion is arrived at on comparing Figs. 4 and 6 with Fig. 5.
In particular, a large space between neighboring phase edges
is helpful to isolate the photonic absorption features
from  each other, and thereby to
localize the columnar features uniformly in the vertical
direction. Therefore, a mid--value of $q$ (i.e., $q\sim 0.5$) is
suggested in order to obtain highly localized and uniform
features.

The formation of
columnar features is due to the spatial characteristics of
propagating Floquet harmonics of the field phasors \c{MEE}. At a  single
frequency, these propagating Floquet harmonics discretely ``beat''
with each other to localize the electric field phasor transversely
(i.e., along the $x$ axis) in the photoresist layer. However, the
electric field phasor has a longitudinal (i.e., along the $z$ axis)
standing--wave profile in the photoresist layer,
because of the spatial attributes of the propagating Floquet
harmonics.  On using broadband sources of illumination, the field
phasors of many different frequencies enter the fray. In other words,
Floquet harmonics of a range of frequencies are generated by the
broadband source to collaboratively produce the total the electric
field in the photoresist layer. Floquet harmonics of different
frequencies correspond to  different classes of both $x$-- and
$z$--variations; and the monochromatic standing--wave
feature appears to be smoothened by the multifrequency Floquet harmonics.
Therefore, the columnar features are  highly localized, and photonic
absorption in the remaining parts of the photoresist layer is highly
uniformized, by broadband illumination. Of course, the
standing--wave pattern would be even less pronounced, especially
near the interface with silicon, by using an ARC that is index--matched  to both the photoresist and silicon for
broadband UV illumination~---~just
as for monochromatic illumination \c{MEE}.

In order to predict the features that might be developed in the
photoresist layer, threshold modeling is often done. For instance,
regions of the RE where the SAR is less than $10\%$ of the maximum SAR are
colored black to bring out the resolved features, while the remaining parts of the
RE are colored white \c{MEE}. This has been done for the three SAR plots shown
in Fig. 10. These were drawn for the same geometry as Figs. 5 and 8,
for the TM illumination mode, and to show the differences between monochromatic
and broadband illumination conditions. Furthermore, the normalization of SAR
means that the plots are independent of the incident power density. 
The exposure wavelengths span the 250--610~nm range
for Fig. 10(a), and the 300--440~nm range for Fig. 10(b), but $\lambda_0=441$~nm
for Fig. 10(c).
Clearly, high--aspect--ratio features of linewidth $<100$ nm are uniformly
well--resolved in the photoresist layer for the full broadband illumination,
but not for the monochromatic illumination. Also, although we have not incorporated the
photochemistry--wavelength response of the photoresist in our calculations, we can conclude that
the standing--wave pattern of features is considerably
diminished by the threshold (development) process for Fig. 10(a),
but not for Fig 10(c)~---~which supports the use of broadband
illumination. Finally, the contrast between Figs.
10(a) and 10(b) implies  the general effectiveness
of the whole spectrum of UV400 illumination in resolving the
patterned features.

Several geometrical factors influence the SAR distribution in the
photoresist layer, and thereby the features developed. One factor
is the period $L$ of the phase--shift mask. Typically, large
values of $L$ are necessary for the formation of stable columnar
features in the photoresist layer \c{MEE}. Figure
11 shows the same SAR distribution in the RE as Fig. 5, but for
the shorter period $L=3$~$\mu$m. Clearly, the columnar features
present in Fig. 11 are not localized as uniformly as those in Fig.
5. In fact, our modeling suggests that $L\geq 4 $ $\mu$m is needed
for the chosen broadband UV400 illumination.

Another influential factor is the thickness $\Delta h_2$ of the
binary phase--shift mask. Experiments indicate that both the resolution and the profile of the
photoresist features after development are very sensitive to the value of $\Delta h_2$ \c{8}, so that
 smallest linewidths are only  achieved at a critical value of
$\Delta h_2\pm10$~nm   on
broadband UV illumination. Leaving aside the thermal and chemical
aspects of NFPSCL, we think that electromagnetic modeling of {\rm
SAR} itself may provide a direct relationship between the groove depth
and the columnar features localized in the photoresist layer. Figure
12 contains the SAR distributions in the RE calculated for
different values of $\Delta h_2$ ranging from 350 nm to 550 nm,
for TM--mode broadband UV400 illumination. Very clearly, the value
of $\Delta h_2$ influences the SAR distribution so significantly
that both highly and uniformly localized columnar features are
present in the photoresist layer only for the restricted range
$\Delta h_2 \in (400, \, 500)$ nm. Hence, the effect of $\Delta
h_2$ on the NFPSCL performance should not be simply viewed as the
phase--shift effect mentioned in Section 1. Instead, it would have
to be understood in a framework that combines the phase--shifting
behavior and the spatial field modulation due to the binary
phase--shift mask \c{Goodman}.

\section{\bf Concluding remarks}
In this paper, we theoretically analyzed photonic absorption  in a photoresist layer employed in
near--field phase--shifting contact lithography on broadband ultraviolet
illumination. The electromagnetic field emitted by  the broadband source
was  represented as a frequency--spectrum of normally
incident plane waves, and a rigorous coupled--wave analysis  was
performed to obtain the absorption spectrum in the photoresist layer.
The specific absorption rate was calculated to characterize the
columnar features localized in the photoresist layer. The narrow
columnar features realized suggest thathigh--aspect--ratio photoresist
features can be printed on the silicon substrate after
development~---~using broadband illumination~---~which is in accord
with experimental observations as exemplified by Fig. 13 and Ref. 8.

Comparison with the results for monochromatic illumination reveals
a dramatic improvement in the resolution and profile of columnar
features by broadband illumination. Furthermore, TM illumination
yields results of higher quality than TE illumination, with the
smallest linewidth less than 100 nm after the threshold
(development) process.

Our calculations indicate that the geometrical dimensions of the
phase--shift mask influence the SAR distribution in the
photoresist layer, and thereby the features developed. In
particular, the shape ratio $q \sim 0.5$, and large values of
period $L\geq 4$ $\mu$m, are needed for obtaining highly and
uniformly localized columnar features in the photoresist layer.
Furthermore, the groove depth $\Delta h_2$ affects the
feature size and profile dramatically, so that restricted values
of $\Delta h_2$ are necessary for the development of
sub--wavelength high--aspect--ratio features by NFPSCL conducted
with broadband UV illumination.

Suppose that $\Delta h_2$ is fixed for a phase shift of $\pi$ at some 
favored wavelength in the central region of the illumination spectrum (see Fig. 3.) The use of an
ARC would definitely assist in the production of desired features by monochromatic
illumination, as is common knowledge in the lithography community. Our electromagnetic
modeling indicates that as good results could be obtained with broadband illumination~---~which
would eliminate the need not only for ARCs but also filtering optics. Although we did not
incorporate here the spectral characteristics of the photochemical response of the photoresist,
our conclusion is buttressed by the numerous experimental examples presented  in Ref. 8
which naturally contained the photochemical effects.

\noindent {\bf Acknowledgements.} The authors acknowledge the
computing assistance provided by Abdul H. Aziz (Institute for High
Performance Computing Applications, Penn State), as well as fruitful
discussions with Martin Peckerar (University of Maryland).  FW thanks the
Penn State Weiss Graduate Program for a Dissertation Fellowship.
This work was also supported in part by the US National Science
Foundation and the US Defense Advanced Research Projects Agency.

\newpage
\begin{figure}[!ht]
\centering \psfull \epsfig{file=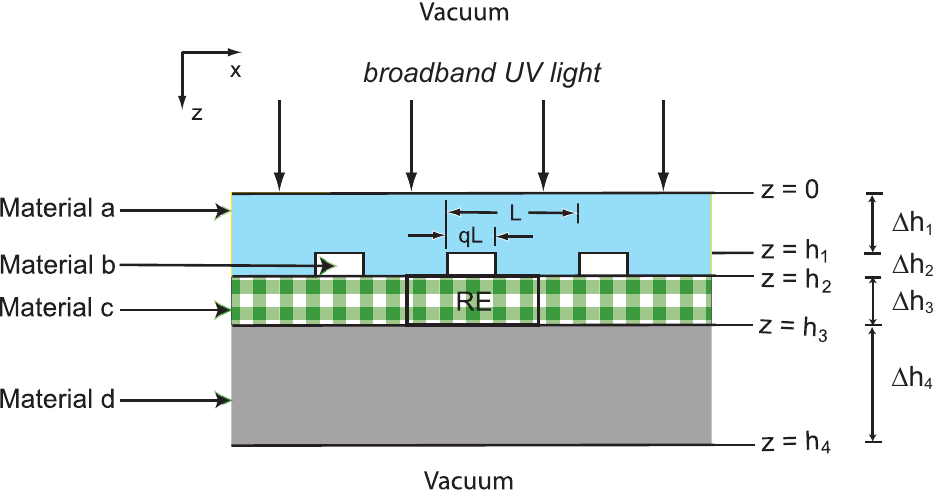,width=5in}
\caption{Schematic of the boundary value problem. SAR
distributions in the region identified as the representative
element (RE)
  are plotted in
Figures 4--12. The length of the RE equals the period $L$, while
its height is $\Delta h_3$.}
\end{figure}

\newpage
\begin{figure}[!ht]
\centering \psfull \epsfig{file=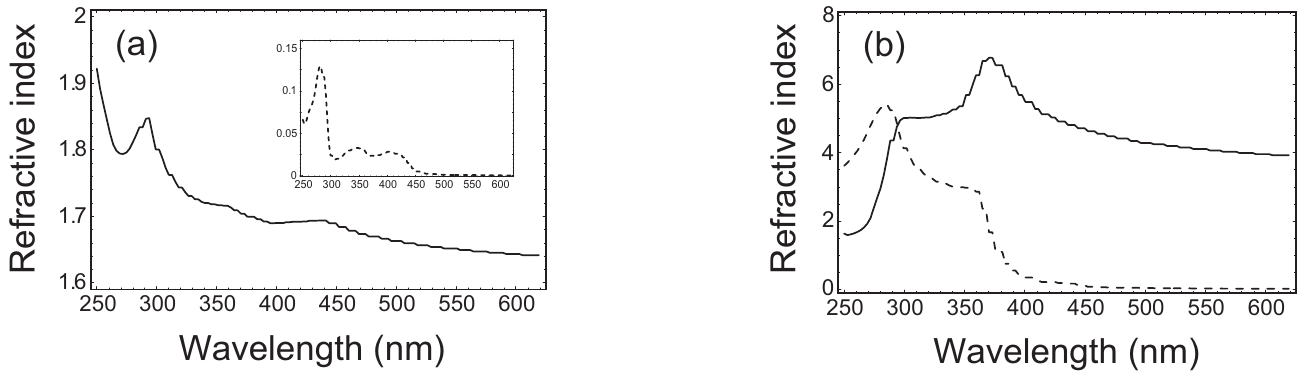,width=5.5in}
\caption{Real (solid lines) and imaginary (dashed lines) parts of the
refractive index   as functions of $\lambda_0 \in [250,\, 610]$
nm for (a) Photoresist SPR 505 and (b) crystalline silicon. The
relative permittivity scalar is the square of the refractive
index.}
\end{figure}

\newpage
\begin{figure}[!ht]
\centering \psfull \epsfig{file=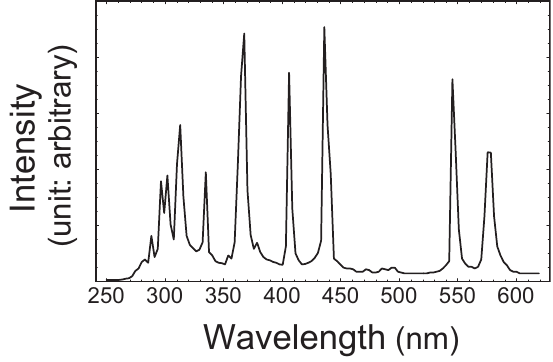,width=3in}
\caption{Spectral intensity $I(\lambda_0)$ of the UV400 source
employed in commercial S\"uss MicroTec mask aligners \c{Suss}. Subsequent filtering
for quasi\-mono\-chromatic transmission at 313, 365 or 435~nm wavelength is often carried out, in standard
lithography practice. The
UV300 source made by the same company has an additional ``dark lens". For all broadband calculations
presented here, the entire spectrum of the UV400 source, as shown in this figure, was used.}
\end{figure}

\newpage
\begin{figure}[!ht]
\centering \psfull \epsfig{file=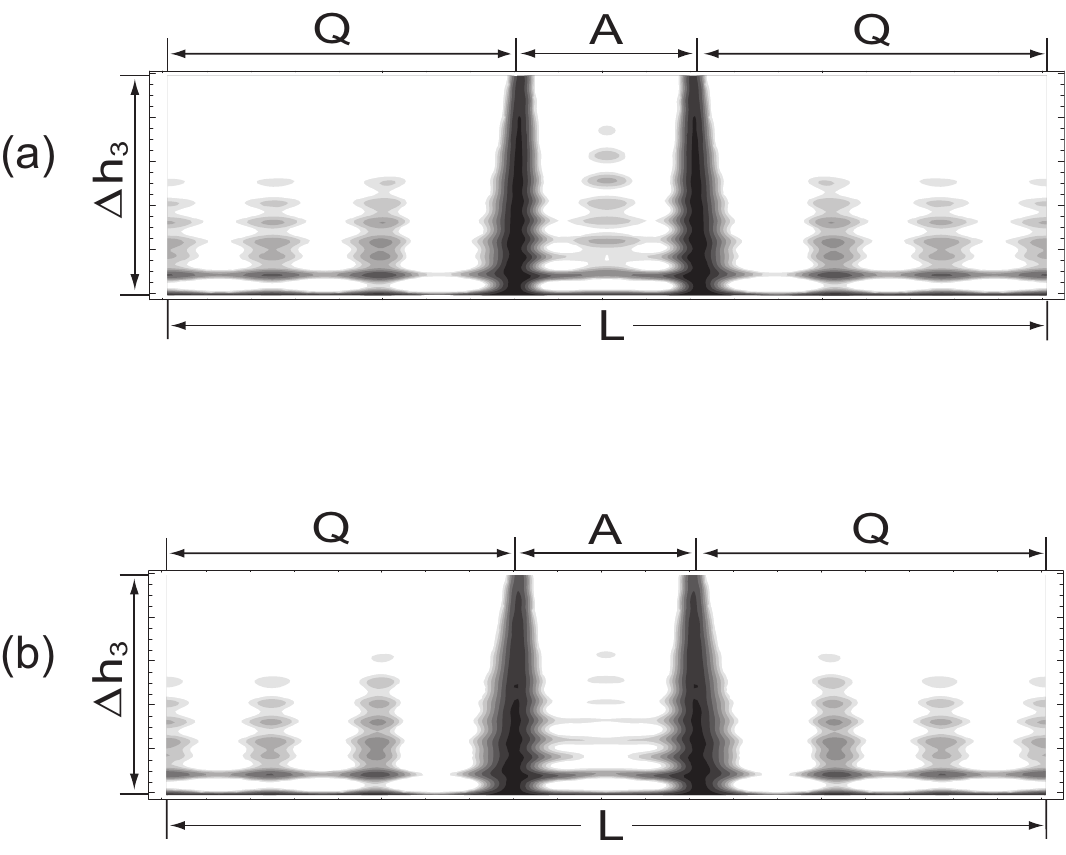,width=4in}
\caption{SAR distribution in the representative element (RE) of
the photoresist layer for broadband illumination ($250\leq \lambda_0\leq 610$~nm) for (a) TE and
(b) TM polarizations. The parameters $\Delta h_2=460$ nm, $\Delta
h_3=1$ $\mu$m, $L=4$  $\mu$m, and $q=0.2$ were employed for
calculation; and a gray--level contour plot of the SAR was
presented by nine gray scales, where black denotes low levels and
white denotes high levels.}
\end{figure}

\newpage
\begin{figure}[!ht]
\centering \psfull \epsfig{file=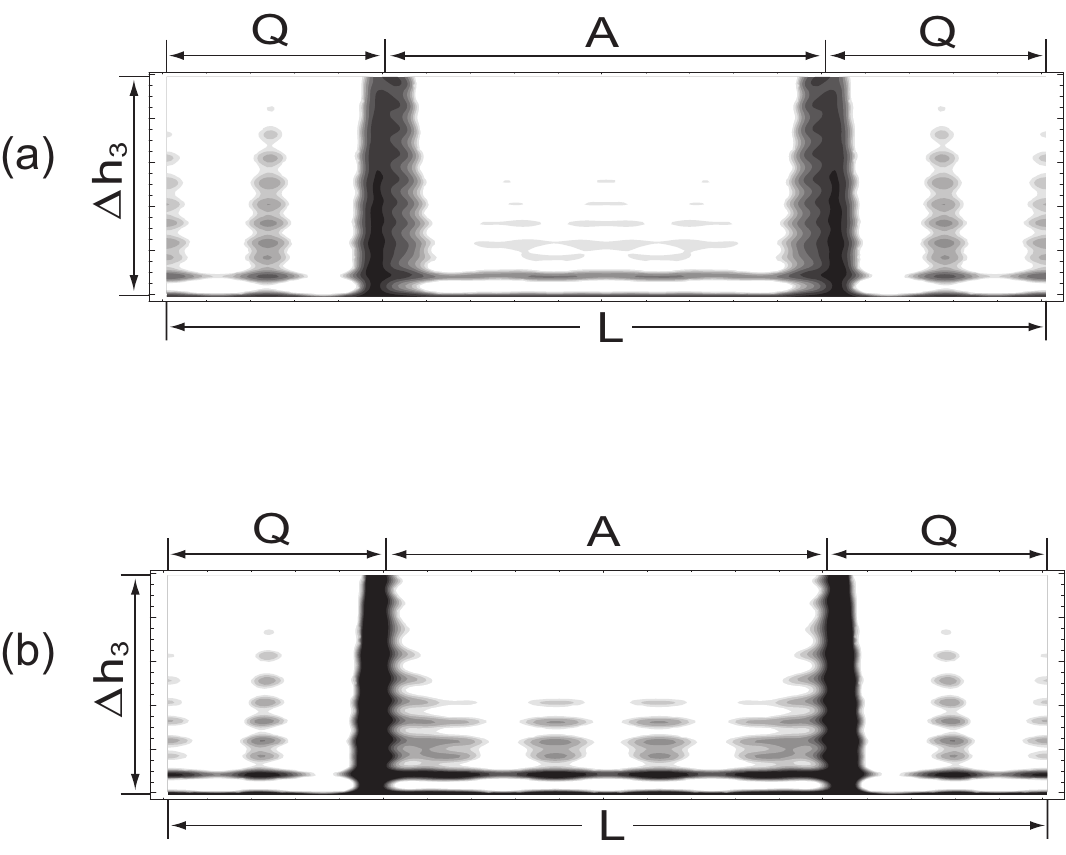,width=4in}
\caption{Same as Figure 4 but for $q=0.5$.}
\end{figure}

\newpage
\begin{figure}[!ht]
\centering \psfull \epsfig{file=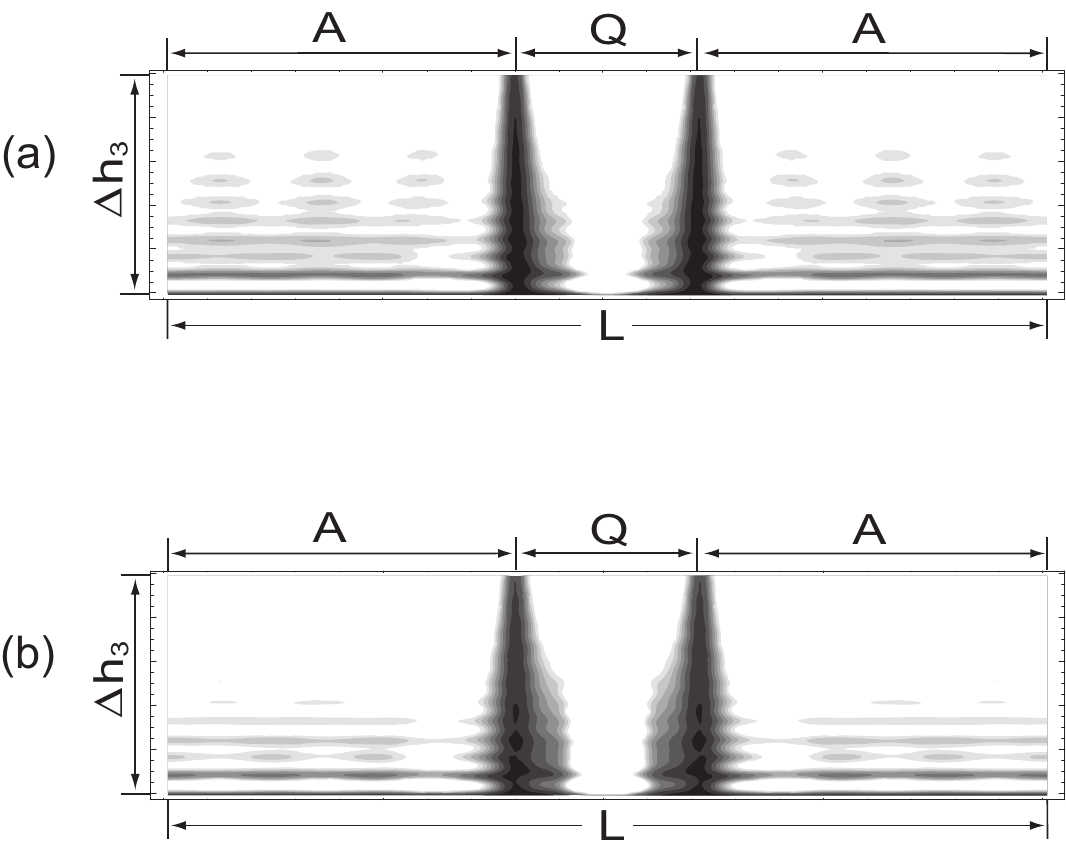,width=4in}
\caption{Same as Figure 4 but for $q=0.8$.}
\end{figure}

\newpage
\begin{figure}[!ht]
\centering \psfull \epsfig{file=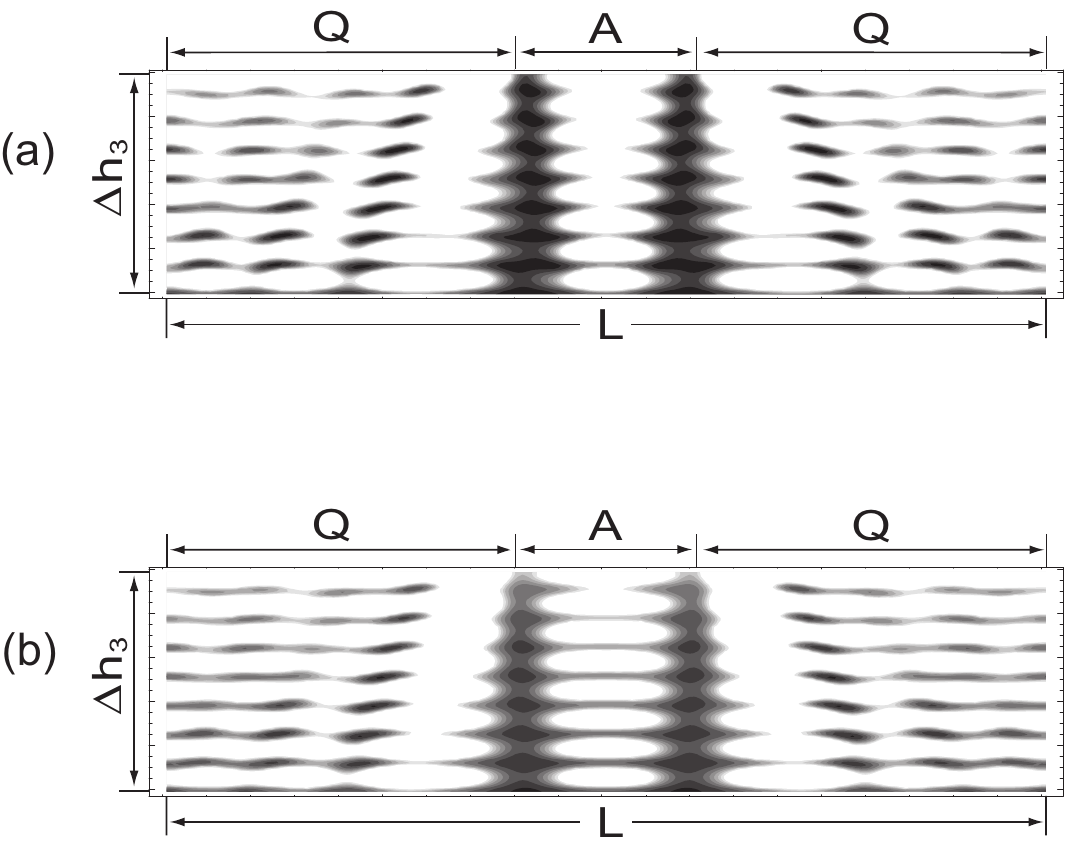,width=4in}
\caption{Same as Figure 4 but for monochromatic illumination at
$\lambda_0=441 $ nm.}
\end{figure}

\newpage
\begin{figure}[!ht]
\centering \psfull \epsfig{file=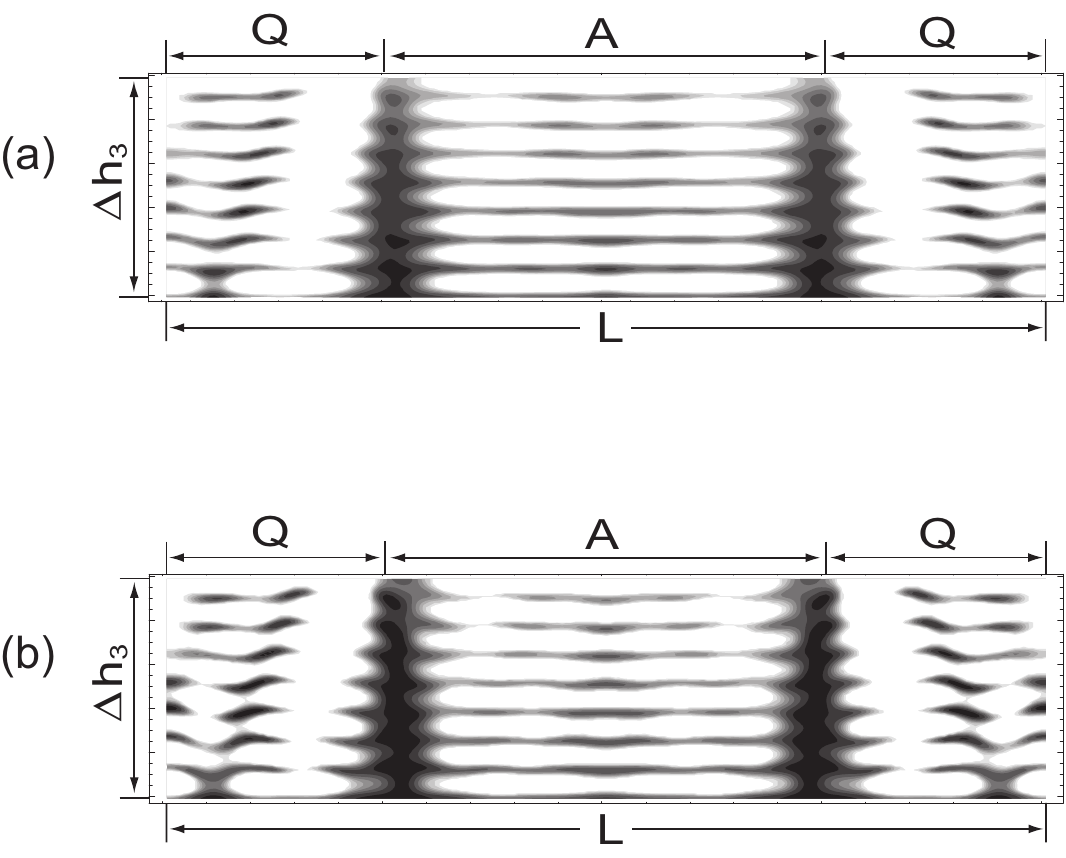,width=4in}
\caption{Same as Figure 7 but for $q=0.5$.}
\end{figure}

\newpage
\begin{figure}[!ht]
\centering \psfull \epsfig{file=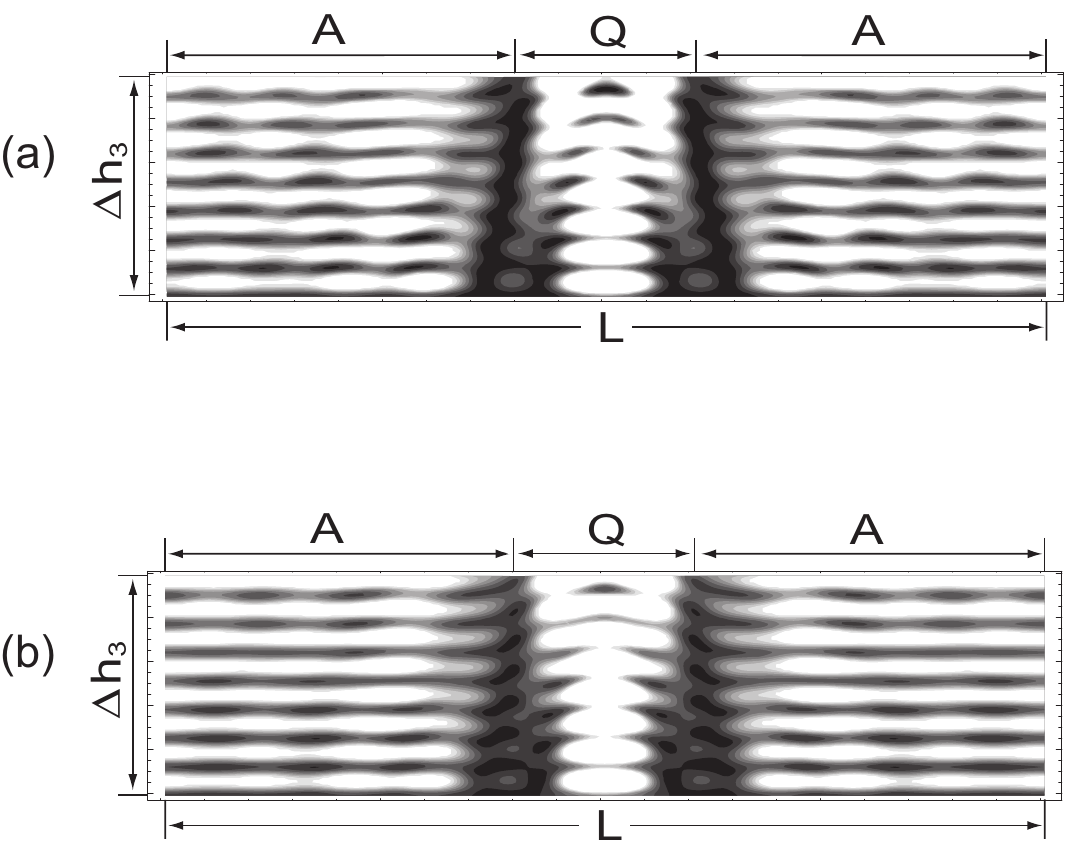,width=4in}
\caption{Same as Figure 7 but for $q=0.8$.}
\end{figure}

\newpage
\begin{figure}[!ht]
\centering \psfull \epsfig{file=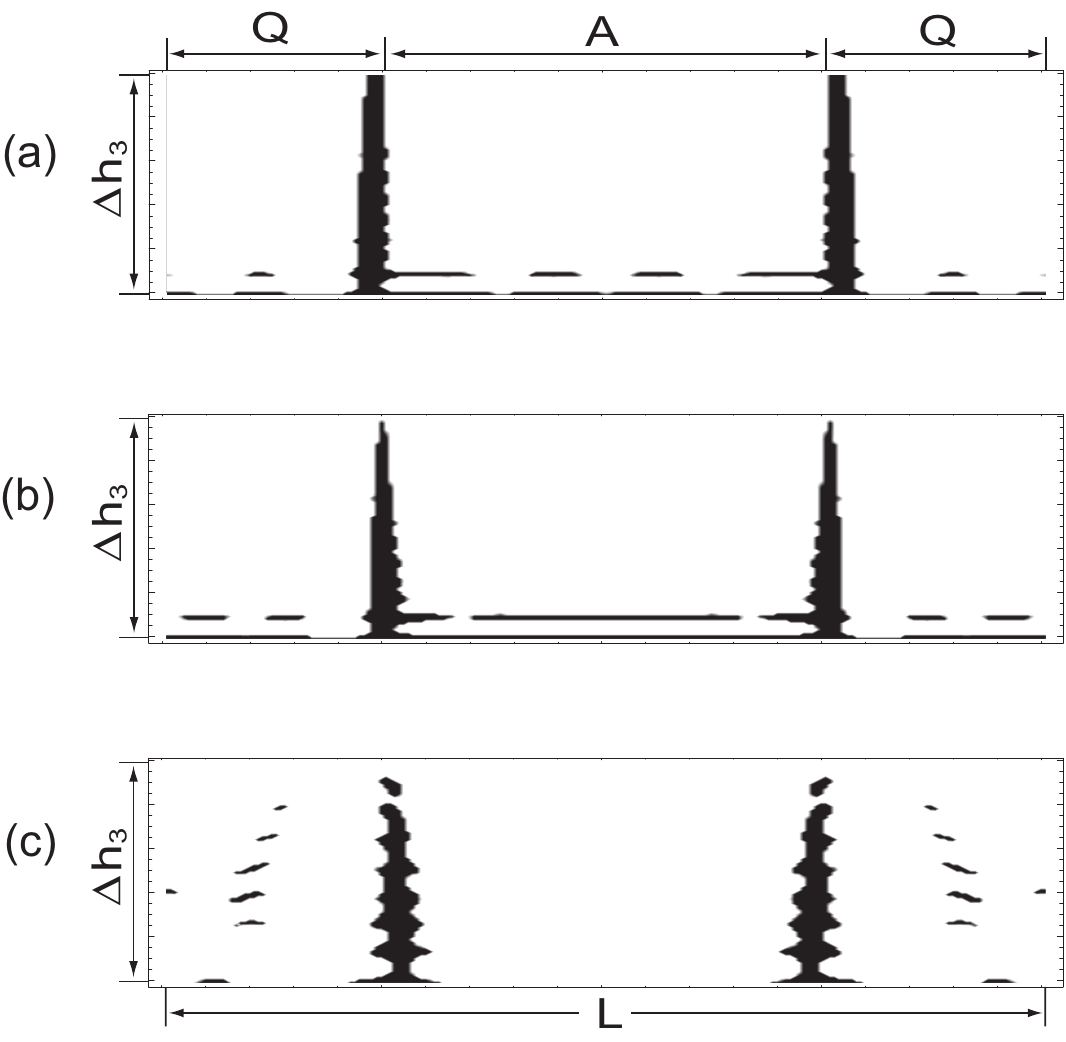,width=4in}
\caption{Post--thresholding SAR plots for TM illumination mode. (a) $250\leq\lambda_0\leq610$~nm,
(b) $300\leq\lambda_0\leq440$~nm, and (c) $\lambda_0=441$~nm.
The remaining parameters are the same as for Figures 5 and 8. Black--colored features correspond
to SAR less than $10\%$ of the maximum SAR, while the white--colored regions are for SAR exceeding
$10\%$ of the maximum SAR.
}
\end{figure}

\newpage
\begin{figure}[!ht]
\centering \psfull \epsfig{file=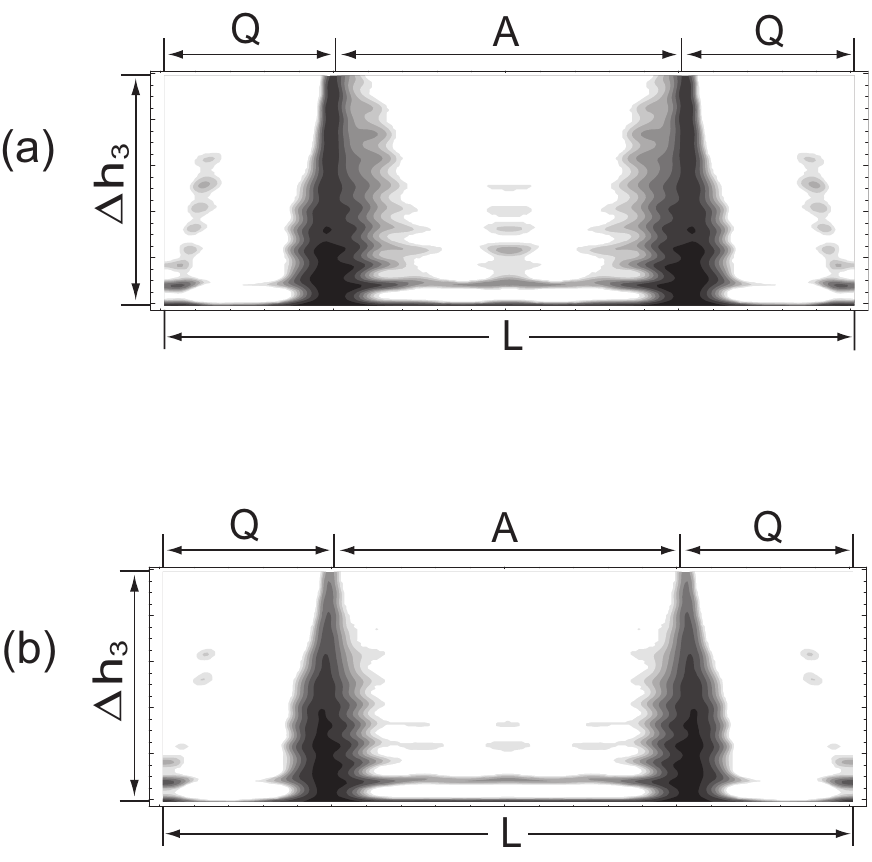,width=3in}
\caption{Same as Figure 5 but for $L=3$ $\mu$m.}
\end{figure}

\newpage
\begin{figure}[!ht]
\centering \psfull \epsfig{file=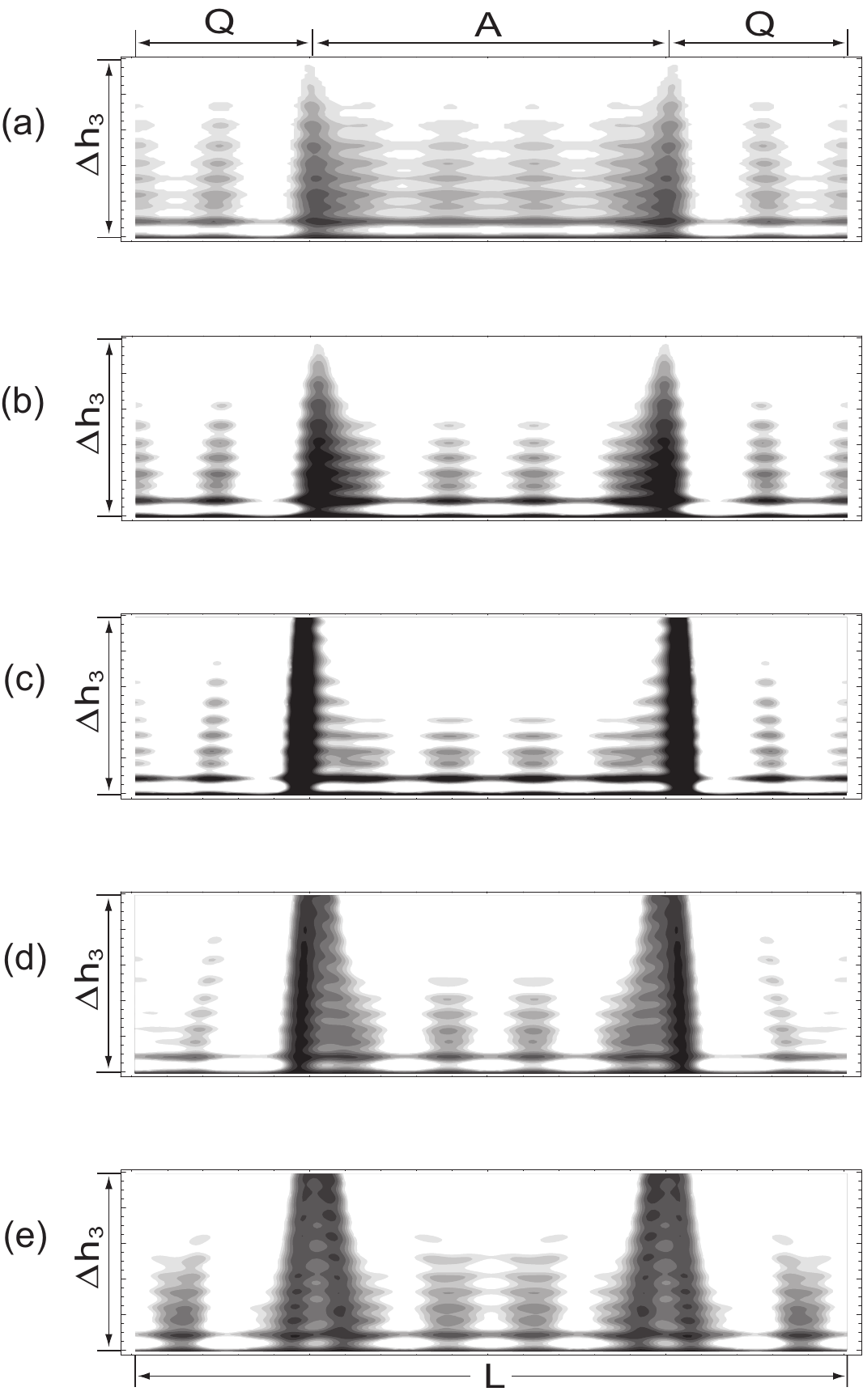,width=4in}
\caption{Same as Figure 5(b) but for different values of $\Delta
h_2$. (a) $\Delta h_2=350 $ nm, (b) $\Delta h_2=400$ nm, (c)
$\Delta h_2=460 $ nm, (d) $\Delta h_2=500 $ nm, and (d) $\Delta
h_2=550 $ nm.}
\end{figure}

\newpage
\begin{figure}[!ht]
\centering \psfull \epsfig{file=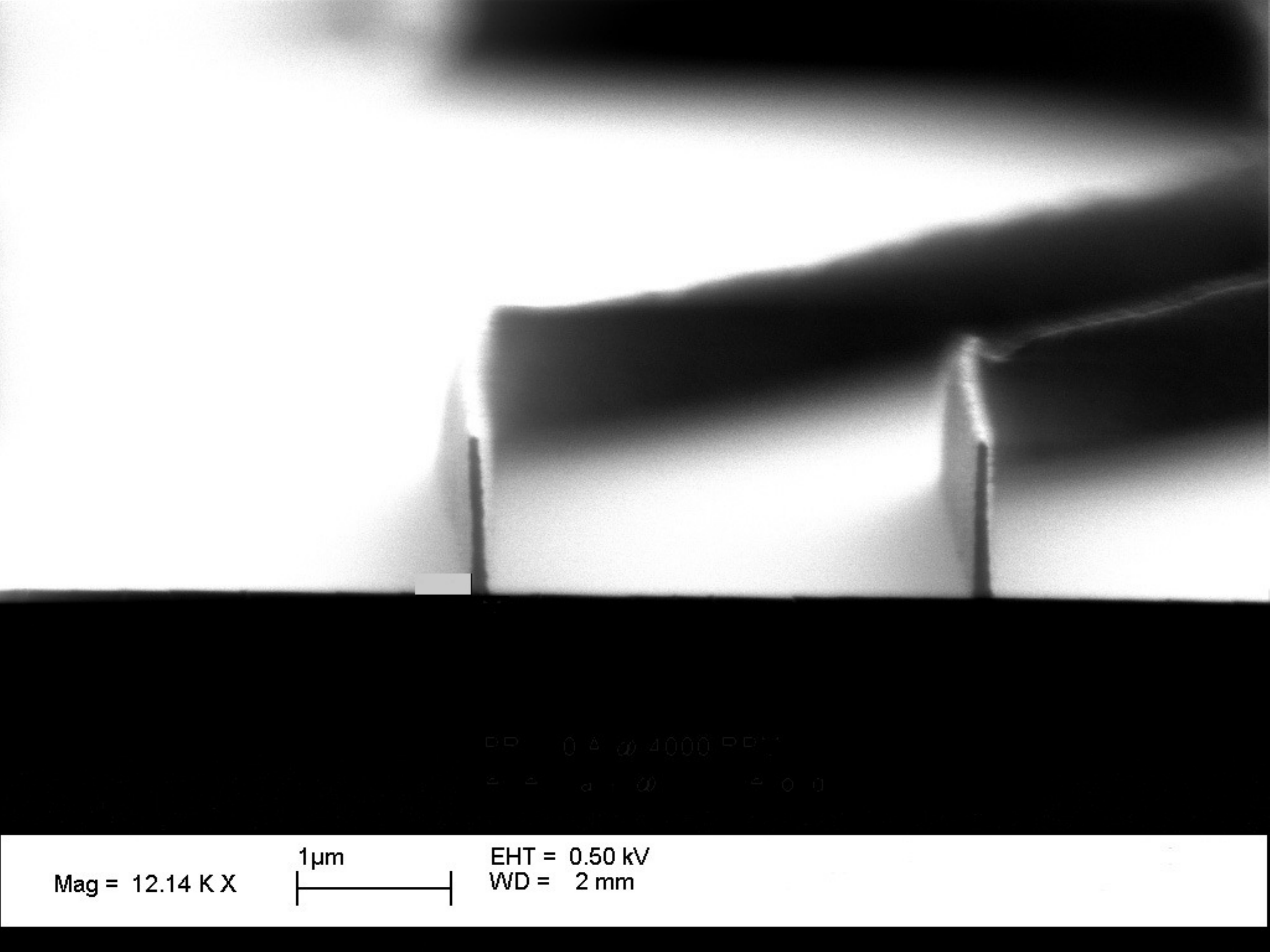,width=4in}
\caption{Two high--aspect--ratio features printed in SPR 510 using a chromeless phase--shifting mask, broadband illumination, and vacuum contact. Exposure
time was 12.5~s, while the UV400 source was used without any filtering optics. }
\end{figure}

\end{document}